\documentclass[journal]{IEEEtran}
\usepackage{amsmath,graphicx}
\begin{document}
\title{Accelerated Evaluation of Automated Vehicles Using Piecewise Mixture Models}

\author{Zhiyuan Huang$^{1}$ , Ding Zhao$^{2}$, Henry Lam$^{1}$, David J. LeBlanc$^{2}$
\thanks{*This work was funded by the Mobility Transformation Center at the University of Michigan with grant no. N021552. Z. Huang and D. Zhao contributed equally to the research.}
\thanks{$^{1}$Zhiyuan Huang ({\tt\small zhyhuang@umich.edu}) and Henry Lam ({\tt\small khlam@umich.edu}) are in the Department of Industrial and Operations Engineering at the University of Michigan.}%
\thanks{$^{2}$Ding Zhao (corresponding author: {\tt\small zhaoding@umich.edu}) and David LeBlanc ({\tt\small leblanc@umich.edu}) are in the University of Michigan Transportation Research Institute.}%
}
\markboth{IEEE TRANSACTIONS ON INTELLIGENT TRANSPORTATION SYSTEMS,~Vol.~XX, No.~XX, December~2016}%
{Shell \MakeLowercase{\textit{et al.}}: Bare Demo of IEEEtran.cls for Journals}

\maketitle

\begin{abstract}
The process to certify highly Automated Vehicles has not yet been defined by any country in the world. Currently, companies test Automated Vehicles on public roads, which is time-consuming and inefficient. We proposed the Accelerated Evaluation concept, which uses a modified statistics of the surrounding vehicles and the Importance Sampling theory to reduce the evaluation time by several orders of magnitude, while ensuring the evaluation results are statistically accurate. In this paper, we further improve the accelerated evaluation concept by using Piecewise Mixture Distribution models, instead of Single Parametric Distribution models. We developed and applied this idea to forward collision control system reacting to vehicles making cut-in lane changes. The behavior of the cut-in vehicles was modeled based on more than 403,581 lane changes collected by the University of Michigan Safety Pilot Model Deployment Program. Simulation results confirm that the accuracy and efficiency of the Piecewise Mixture Distribution method outperformed single parametric distribution methods in accuracy and efficiency, and accelerated the evaluation process by almost four orders of magnitude.  \\
\end{abstract}


\IEEEpeerreviewmaketitle

\section{Introduction}
\IEEEPARstart{I}{t} is critical to thoroughly and rigorously test and evaluate an Automated Vehicle (AV) before its release. Recent crashes involving a Google self-driving car \cite{GoogleAutoLLC} and a Tesla Autopilot vehicle \cite{PreliminaryHWY16FH018} attracted the public's attention to AV testing and evaluation. While these AVs are generally considered as industrial leaders, because they use public road for testing, statistically they have not yet accumulated enough miles.  The Tesla Autopilot, in particular, was criticized for being released too early in the hands of the general public \cite{Evan2016FatalPerfect}. 

Currently, there are no standards or protocols to test AVs at automation level 2 or higher. Many companies adopt the Naturalistic Field Operational Tests (N-FOT) approach \cite{FESTA-Consortium2008}. However, this method is inefficient because safety critical scenarios rarely happen in daily driving. The Google Self-driving cars accumulated 1.9 million driving. This distance, although sounds a lot, provides limited exposure to critical events, given that U.S. drivers encounter a police reported crash every five hundred thousand miles on average and fatal crash every one hundred million miles \cite{NHTSA2014b}. In the meantime, both Google and Tesla update their software throughout the process, which may have improved safety, but the newest version of the AV has not accumulated that many miles as they have claimed. In summary, today's best practice adopted by the industry is time-consuming and inefficient.  A better approach is needed. 

\subsection{Related Researches}

Besides the N-FOT, the test matrix approach \cite{Peng2012EvaluationVehicles,Aust2012EvaluationSystems} and the worst-case scenarios approach \cite{Kou2010DevelopmentSystems,Ungoren2005AnModel,Ma1999ASystems} are two alternative methods for vehicle evaluation. Our approach follows the Accelerated Evaluation concept we proposed \cite{Zhao2016AcceleratedTechniques} to provide a brand-new alternative. The basic concept is that as high-level AVs just began to penetrate the market, they mainly interact with human-controlled vehicles (HVs). Therefore we focus on modeling the interaction between the AV and the HV around it.  The evaluation procedure involves four steps:
\begin{itemize}
	\item Model the behaviors of the “primary other vehicles” (POVs) represented by $f(x)$ as the major disturbance to the AV using large-scale naturalistic driving data
	\item Skew the disturbance statistics from $f(x)$ to modified statistics $f^{*}(x)$ (accelerated distribution) to generate more frequent and intense interactions between AVs and POVs
	\item Conduct “accelerated tests” with $f^{*}(x)$ 
	\item Use the Importance Sampling (IS) theory to “skew back” the results to understand real-world behavior and safety benefits
\end{itemize}

\begin{figure}[t]
      \centering
   \includegraphics[width=\linewidth]{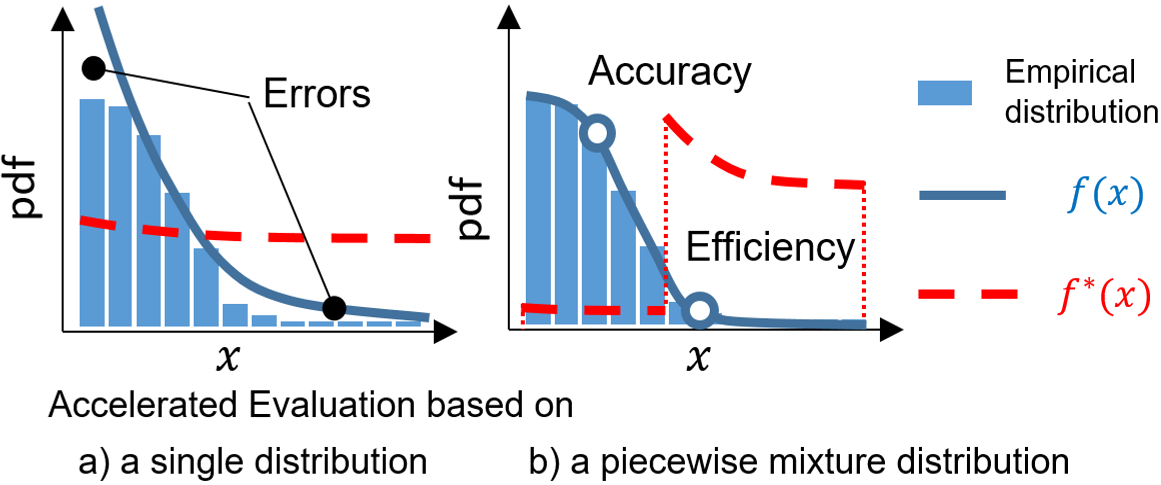}
      \caption{Acceleration evaluation based on single parametric distribution and Piecewise Mixture Distribution.}
      \label{fig:g1}
\end{figure}

This approach has been successfully applied to evaluate AVs in the frontal crash with a cut-in vehicle \cite{Zhao2016AcceleratedTechniques} and also frontal crash with a lead vehicle \cite{Zhao2015j,Zhao2016g}. This approach was confirmed to significantly reduce the evaluation time while accurately preserving the statistical behavior of the AV-HV interaction. In the previous studies, the evaluation time was reduced by two to five orders of magnitudes -  the accelerated rate depends on the test scenarios, where rarer events achieve higher accelerated rate.  The non-accelerated models and the accelerated models were built based on signal component distributions. While this method does benefit from its simple mathematical form, it has a few drawbacks as illustrated in Fig. \ref{fig:g1}. i) The fitting of the rare events (usually the tail part of the statistical distributions) would be dominated by the fitting of the normal driving behaviors (the majority part of the distributions), which may induce large errors. ii) The full potential in higher accelerated rate is not achieved due to the lack of flexibility of the modified accelerated models. 

\subsection{Contribution}
In this paper, we proposed a more general framework for the Accelerated Evaluation method to overcome the aforementioned limitations based on Piecewise Mixture Distribution Models as illustrated in Fig. \ref{fig:g1} b). In this paper, we implemented the Accelerated Evaluation method under the new framework. Comparing to our previous work \cite{Huang2016UsingScenario}, we thoroughly discuss the Cross Entropy method with proposed framework in this paper. We present practical tips to overcome numerical issues and reduce computational efforts. We demonstrate this method by evaluating the longitudinal control system reacting to vehicles making cut-in lane changes. 
%
%
%

\subsection{Paper Structure}
Section \ref{sec:single} will introduce the lane change model based on single parametric distributions. In Section \ref{mixture_section}, we present the new lane change model with Piecewise Mixture Distributions. We establish the Accelerated Evaluation in Section \ref{sec:accelerated_eval} and discuss the Cross Entropy method with Piecewise Mixture Distribution models in Section \ref{sec:ce}. Simulation results are discussed in Section \ref{sec_simulation}. Section \ref{sec:conclusions} concludes this paper.

\section{Accelerated Evaluation with Single Parametric Distributions}
\label{sec:single}
The lane change events were extracted from the Safety Pilot Model Deployment (SPMD) database \cite{Bezzina2014}.  With over 2 million miles of vehicle driving data collected from 98 cars over 3 years, we identify 403,581 lane change events.  Previously \cite{Zhao2016AcceleratedTechniques}, we used 173,692 events with a negative range rate to build a statistical model focusing on three key variables that captured the effects of gap acceptance of the lane changing vehicle: velocity of the lead vehicle ($v_L$), range to the lead vehicle ($R_L$) and time to collision ($TTC_L$). $TTC_L$ was defined as:
\begin{equation}
	TTC_L=- \frac{R_L}{\dot{R_L}},
\end{equation}
where $\dot{R_L}$ is the relative speed. 

The modeling of these three variables was hard to handle because of dependency, so we simplified it based on a crucial observation. Although $TTC_L$ is dependent on $v_L$ generally, we split the data into 3 segments: $v_L$ at 5 to 15 m/s, 15 to 25 m/s and 25 to 35 m/s. Within each segment, $R_L$ is independent with $v_L$ and $TTC_L$. This allowed us to model $TTC_L$ and $R_L$ independently with regard to the value of $v_L$. By comparing among 17 types of commonly used distribution templates, we selected the Pareto distribution to model $R_L^{-1}$ and used the exponential distribution for $TTC_L^{-1}$ segments.

\begin{figure}[t]
      \centering
   \includegraphics[width=\linewidth]{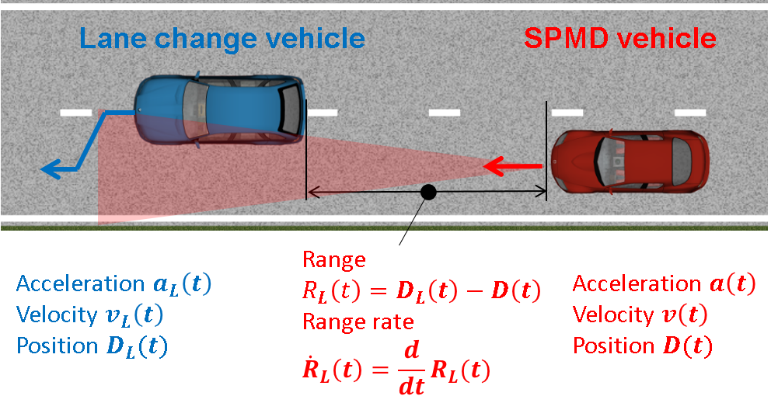}
      \caption{Lane change data collected by SPMD vehicle.}
      \label{fig:o1}
\end{figure}

Using the empirical distribution of $v_L$ and parametric distributions of $R_L$ and $TTC_L$, we drew values from these distributions as inputs to simulate the AV-HV interaction. The outcome from the simulation can be considered as an event indicator function $I_\varepsilon(x)$ that returns $\{1, 0\}$ depending on the event of interest. Given the stochastic distribution of the variables and the event indicator function, we obtained the optimal exponential distribution for Importance Sampling by implementing the Cross Entropy method \cite{rubinstein1999rare}. As we have shown in Fig. \ref{fig:g1} a), we used only single parametric distributions. In the next section, we introduce our new approach using Piecewise Mixture Distributions.

\section{Lane Change Model with Piecewise Mixture Distributions }
\label{mixture_section}

Although many commonly used parametric distributions have concise and elegant forms, they do not always describe the data distribution well. Instead, a better fitting can be achieved by dividing the dataset into several subsets. We estimate the model parameters using the Maximum Likelihood Estimation (MLE) \cite{Aldrich1997RA1912-1922} in each subset. The general process of MLE is as follow.
%
%
%

Assume we have a family of distribution with Cumulative Distribution Function (CDF) $F(x|\theta)$, where $\theta$ is the parameter vector of $F$. The corresponding Probability Density Function (PDF) of $F$ is $f(x|\theta)$. Assuming that data $D=\{X_1, X_2,...,X_N\}$ is independently and identically distributed and the distribution is in the family of $F(x|\theta)$, we want to find the most ``likely" parameter $\hat{\theta}$.

We define the likelihood function \cite{Cox1979TheoreticalStatistics} as \begin{equation}
	L(\theta|D)=P(D|\theta)=\Pi_{n=1}^{N} f(X_n|\theta).
\end{equation}
We call the estimation of $\hat{\theta}$ that maximizes the likelihood function the mostly likely estimation MLE.
%
%
%

For computation convenience, we introduce the log-likelihood function
    \begin{equation}
		\mathcal{L} (\theta|D)=\ln L(\theta|D)=\sum_{n=1}^{N} \ln f(X_n|\theta).
	\end{equation}	
Since the logarithm is monotone, the log-likelihood function preserves the optimizer of the original function. \cite{Boyd2004ConvexOptimization} The optimizer of log-likelihood function, $\hat{\theta}$, is the MLE of distribution family $F$. We have the MLE as \begin{equation}
\hat{\theta} = \arg \max_{\theta}  \  \mathcal{L}(\theta|D).
\end{equation}

In the following, we describe the Piecewise Mixture Distribution fitting concept based on MLE and we present the bounded distribution fitting results. All optimization problems presented in this section are tractable and can be solved by {\bf fminunc} in MATLAB.

	\subsection{General Framework of the Piecewise Mixture Distribution Lane Change Model}
We define Piecewise Mixture Distribution to be distribution with CDF in the form of 
\begin{equation}
\label{eq:CDF}
		F(x)=\sum_{i=1}^{k} \pi_i F_{\theta_i}(x |\gamma_{i-1} \leq x< \gamma_i),
\end{equation}
where $k$ is the number of truncation, $\sum_{i=1}^{k}  \pi_i=1$, and $F_i(x |\gamma_{i-1} \leq x< \gamma_i)$ is the conditional cumulative distribution function, meaning that $F_i(\gamma_{i-1} |\gamma_{i-1} \leq x< \gamma_i)=0$ and $F_i(\gamma_i |\gamma_{i-1} \leq x< \gamma_i)=1$ for $i=1,...,k$. $\theta_i$ denotes the parameter(s) for $F_i$. We can consider that $\pi_i=P(\gamma_{i-1} \leq x< \gamma_i)$ and when $x \geq 0$, we have $\gamma_0=0$ and $\gamma_k=\infty$. By this definition, the PDF of the Piecewise Mixture Distribution is \begin{equation}
	f(x)=\sum_{i=1}^{k} \pi_i f_{\theta_i}(x |\gamma_{i-1} \leq x< \gamma_i).
\end{equation}

	In our case, $\theta= \{\pi_1,...,\pi_k, \theta_1,...,\theta_k\}$. Splitting $D$ into pieces regarding the truncation points $\{\gamma_1,...,\gamma_{k-1}\}$, gives data index sets $S_i=\{j|\gamma_{i-1} \leq X_j<\gamma_i\}$ for $i=1,...,k$. We can write the log-likelihood function as
    \begin{equation}
    \label{eq:mle}
    	\begin{array}{l}
		\mathcal{L}(\theta|D)  =\sum_{i=1}^{k}     \sum_{n \in S_i} \ln \pi_i	\\
      \hspace{3em} +\sum_{i=1}^{k}     \sum_{n \in S_i} \ln f_{\theta_i}(X_n  | \gamma_{i-1} \leq x< \gamma_i) .
        \end{array}
	\end{equation}
    
	We obtain the MLE of $\theta$ can be obtained by maximizing $\mathcal{L}(\theta|D)$ over $\theta$. Since $\mathcal{L}$ is concave over $\pi_i$, we take 
\begin{equation}
	\frac{\partial \mathcal{L}}{\partial \pi_i} =0
\end{equation}
and get \begin{equation}
	\hat{\pi}_i= |S_i|/N. 
\end{equation}

	For parameters $\theta_i$ in $F_i$, it is known (\ref{eq:mle}) to be the same as computing the MLE of $\theta_i$ with corresponding dataset $D_i=\{X| \gamma_{i-1} \leq X< \gamma_i  \ and \ X \in D\}$. Since we use bounded distribution for each $F_i$, below we explain the estimation of parameters for the three distributions we applied in later sections.
    
   	To sample from a Piecewise Mixture Distribution, we could use the inverse function approach. See Appendix \ref{apped_inv} for the details.
	
	\subsection{Bounded Distribution}
We develop three bounded distributions and use them in the lane change model.
	
	\subsubsection{MLE for bounded exponential distribution}
	
The bounded exponential distribution with rate $\theta$ has the form\begin{equation}
	f(x|\gamma_1 \leq x < \gamma_2) = \frac{\theta e ^{-\theta x}}{e^{-\theta \gamma_1}-e ^{-\theta \gamma_2}}
\end{equation}
for $\gamma_1 \leq x < \gamma_2$. 
    
    For dataset $D=\{X_1,...,X_N\}$, the log-likelihood function is\begin{equation}
	\mathcal{L}(D|\theta)=\sum_{n=1}^{N} \ln \theta -\theta X_n - \ln(e ^{-\theta \gamma_1}-e ^{-\theta \gamma_2}),
%
%
%
\end{equation}
where $\mathcal{L}$ is concave over $\theta$. Although we cannot solve the maximization analytically, it is solvable through numerical methods.
	
	Therefore, the MLE of $\theta$ is given by the optimization\begin{equation}
	\max_\theta  \  N\ln \theta - N \ln(e ^{-\theta \gamma_1}-e ^{-\theta \gamma_2})-\sum_{n=1}^{N}\theta X_n.
\end{equation}

	\subsubsection{MLE for bounded normal distribution}
	
	Consider a bounded normal distribution with mean 0 and variance $\theta^2$ conditional on $0 \leq \gamma_1 \leq x < \gamma_2 $. The PDF is \begin{equation}
	f(x|\gamma_1 \leq x < \gamma_2)= \frac{\frac{1}{\theta}\phi(\frac{x}{\theta})}{\Phi(\frac{\gamma_2}{\theta})-\Phi(\frac{\gamma_1}{\theta})}.
\end{equation}
	
	The MLE of the bounded normal distribution is given by

\begin{equation}
\max_\theta -\frac{\sum_{n=1}^{N} X_n^2}{2\theta^2}-N\ln\theta-N\ln(\Phi(\frac{\gamma_2}{\theta})-\Phi(\frac{\gamma_1}{\theta})).
\end{equation}

\subsubsection{Fitting mixture model with EM algorithm}
Compared to single parametric distributions, mixture distribution combines several classes of distribution and thus is more flexible. We consider the fitting problem of mixture bounded normal distribution.

The PDF of mixture of $m$ bounded normal distribution can be written as\begin{equation}
f(x|\gamma_1 \leq x < \gamma_2)=\sum_{j=1}^{m} p_j f_j(x|\gamma_1 \leq x < \gamma_2)
\end{equation}
where $f_j$ is bounded Gaussian distribution with mean 0 and variance $\sigma_j^2$. The parameters here are $\theta=\{p_1,...,p_m,\sigma_1^2,...,\sigma_m^2\}$. We want to find MLE of $p_j$ and $\sigma_j^2$ for $j=1,...,m$.

The log-likelihood function for data $D=\{X_n\}_{n=1}^N$ is\begin{equation}
\mathcal{L}(\theta|D)=\sum_{n=1}^{N} \ln  \sum_{j=1}^{m} p_j f_j(X_n|\gamma_1 \leq x < \gamma_2).
\end{equation}
We note that this is hard to solve directly, because there is a sum within the log function. Therefore, we apply the Expectation-Maximization (EM) \cite{Dempster1977b} algorithm to find the optimizer, i.e. MLE, for the parameters.

We define $Z_n^j$ to denote whether or not the random number $X_n$ comes from mixture distribution $j$, $j=1,...,m$, and $Z_n^j=\{0, 1\}$. We also introduce the expectation\begin{equation}
E[Z^j_n|X_n]:=\tau_n^j.
\end{equation}

The EM algorithm starts with initial parameters $\{p_j,\sigma_j\}$, $j=1,...,m$. For data $D=\{X_n\}_{n=1}^N$, we set complete data as $D_c=\{X_n, Z_n\}_{n=1}^N$. The EM algorithm optimizes $E[\mathcal{L}(\theta|D_c)|D]$ in every step. The E step updates $E[\mathcal{L}(\theta|D_c)|D]$, and the M step optimizes this function. The algorithm iterates E step and M step until reaching the convergence criterion.

In our case, \begin{equation}
\begin{array}{l}
E[\mathcal{L}(\theta|D_c)|D]=\sum_{n=1}^{N} \sum_{j=1}^{m} \tau_n^j \left(\ln p_j + \ln f_j(X_n)\right). 
\end{array}
\end{equation}

Since objective $E[l_c(\theta|D_c)|D]$ in the M step is concave over $p_j$ and $\sigma_j$, we could maximize the objective function through an analytic approach for $p_j$:	\begin{equation}
	p_j=\frac{\sum_{n=1}^{N} \tau_n^j }{N}.
\end{equation}
For $\sigma_j$, we can solve the following maximization problem through numerical approach.
\begin{multline}
	\sigma_j=\arg \min_{\sigma_j} -\tau_n^j \ln \sigma_j+ \tau_n^j \ln \phi\left(\frac{X_n}{\sigma_j}\right)-\\ \tau_n^j \ln \left(\Phi(\frac{\gamma_2}{\sigma_j})-\Phi(\frac{\gamma_1}{\sigma_j})\right).
\end{multline}

See Appendix \ref{apped_em} for the full EM algorithm.

\section{Accelerated Evaluation with Importance Sampling}
\label{sec:accelerated_eval}

Importance Sampling (IS) is thus used to accelerate the evaluation process, because crude Monte Carlo simulations for rare events can be time-consuming. Below we describe the IS method. 

\subsection{Important Sampling and Optimal IS distribution}
Let $x$ be a random variable generated from distribution $F$, and $\varepsilon \subset \Omega$ where $\varepsilon$ is the rare event of interest and $\Omega$ is the sample space. Our objective is to estimate \begin{equation}
{P}(X \in \varepsilon)=E[I_\varepsilon(X)]=\int I_\varepsilon(x) dF
\end{equation}
where\begin{equation}
	I_\varepsilon(x)=\begin{cases} 1 & x \in \varepsilon,\\
0 & otherwise.\end{cases}
\end{equation}

We can write the evaluation of rare events as the sample mean of $I_\varepsilon(x)$ \begin{equation}
	\hat{P}(X \in \varepsilon) = \frac{1}{N} \sum_{n=1}^N I_\varepsilon(X_n),
\end{equation}
where $X_i$'s are drawn from distribution $F$.

Since we have \begin{equation}
	E[I_\varepsilon(X)]=\int I_\varepsilon(x) dF = \int I_\varepsilon(x) \frac{dF}{dF^*} dF^* ,
\end{equation}
we can compute the sample mean of $I_\varepsilon(X) \frac{dF}{dF^*}$ over the distribution $F^*$, which has the same support with $F$, to obtain an unbiased estimation of ${P}(X \in \varepsilon) $. By appropriately selecting $F^*$, the evaluation procedure obtains an estimation with smaller variance. This is known as Importance Sampling \cite{Bucklew2004a} and $F^*$ is the IS distribution.

For estimating ${P}(X \in \varepsilon)$,we note that an optimal IS distribution\begin{equation}
	F^{**}(x)=F(x|\varepsilon)=\frac{P(X\leq x,\ \varepsilon)}{P(x\in \varepsilon)}
    \label{eq:zerovar}
\end{equation} could reduce the variance of IS estimation to 0, but the optimal requires the knowledge of ${P}(X \in \varepsilon)$. However, it guides the selection of the IS distribution.

\subsection{Exponential Change of Measure}
Exponential change of measure is commonly used to construct $F^*$. Although the exponential change of measure cannot guarantee convergence to optimal distribution, it is easy to implement and the new distribution generally stays within the same class of distribution.

Exponential change of measure distribution takes the form of \begin{equation}
f_\theta(x)=\exp( \theta x - \kappa (\theta))f(x),
%
%
%
\end{equation}
where $\theta$ is the change of measure parameter and $\kappa(\theta)$ is the log-moment generating function of original distribution $f$. When $\theta=0$, we have $f_\theta(x)=f(x)$.

For a bounded exponential distribution, the exponential change of measure distribution is \begin{equation}
	f_{\theta}(x| \gamma_1 \leq x < \gamma_2)=\frac{(\lambda-\theta)e^{-(\lambda-\theta)x}}{e^{-(\lambda-\theta)\gamma_1}-e^{-(\lambda-\theta)\gamma_2}},
\end{equation} where $\lambda$ is the parameter for exponential distribution. We note that $f_\theta$ is still a bounded exponential distribution and $\lambda_\theta=\lambda-\theta$.

For a bounded normal distribution, the exponential change of measure distribution is \begin{equation}
	f_\theta(x| \gamma_1 \leq x < \gamma_2)=\frac{\frac{1}{\sigma}\phi(\frac{x-\sigma^2 \theta}{\sigma})}{\Phi(\frac{\gamma_2-\theta \sigma^2}{\sigma})-\Phi(\frac{\gamma_1-\theta \sigma^2}{\sigma})},
\end{equation} where the original distribution truncated from a normal distribution with parameters $\mu=0$ and $\sigma$. We note that the change of measure distribution is still a bounded normal distribution with $\mu=\theta \sigma^2$ and $\sigma$.

\section{Cross Entropy Method and Implementation}
\label{sec:ce}

Section \ref{sec:accelerated_eval} discussed optimal IS distribution $F^{**}$ providing 0 variance estimation to the value of interest, whereas this section describes the Cross Entropy method used to estimate the ``optimal" parameters $\theta$, which minimizes the ``distance" between a parametric distribution $F_\theta$ and $F^{**}$ without knowing $F^{**}$. The description below is based on the Piecewise Mixture Distribution structure.
\subsection{Introduction}

The Cross Entropy, which is also known as Kullback-Leibler distance \cite{Vapnik1998StatisticalTheory}, measures the similarity between distributions. We define the Cross Entropy between function g and h as\begin{multline}
\mathcal{D} (g,h)=E_g [ln \frac{g(X)}{f(X)}]=\int g(x)\ln g(x) dx -\\ \int g(x)\ln h(x) dx .
\end{multline} 

From (\ref{eq:zerovar}), we know that the PDF of the optimal IS distribution $F^{**}$ is \begin{equation}
	f^{**}(x)=\frac{I_\varepsilon(x)f(x)}{{P}(x \in \varepsilon)} .
\end{equation}

Since ${P}(x \in \varepsilon)$ is generally unavailable, we use a parametric distribution $F_\theta$ to approach the optimal IS distribution. We want to find the parameter $\theta^*$ that minimizes the Cross Entropy \cite{Kroese2013a} between $f^{**}$ and $f_\theta$. We denote $\theta^*$ as the optimal parameter for the parametric distribution. Then the minimization problem \begin{equation}
\min_{\theta} \mathcal{D} (f_{\theta},f^{**}) 
\end{equation}
is equivalent to\begin{equation}
	\max_\theta \ E_{\theta_s}[I_\varepsilon(X) \frac{f(X)}{f_{\theta_s}(X)} \ln f_\theta(X)],
    \label{eq:CEobj}
\end{equation}
where $f_{\theta_s}$ denotes the sampling distribution with parameters $\theta_s$. We note that this is a generalized setting, since we can use any sampling distribution $f_{\theta_s}$ as long as it has the same support with $f$. This is the baseline for iterations in the Cross Entropy method. We use the same form as $f_\theta$ because in the following sections, we use a sampling distribution which is in the same family as the parametric distribution.

We estimate $\theta^*$ by solving the stochastic counterpart of (\ref{eq:CEobj})\begin{equation}
	\max_\theta \ \frac{1}{N} \sum_{n=1}^{N} I_\varepsilon(X_n) \frac{f(X_n)}{f_{\theta_s}(X_i)} \ln f_\theta(X_n),
    \label{eq:sto_obj}
\end{equation}
where samples $\{X_1,...,X_N\}$ are drawn from the sampling distribution $f_{\theta_s}$.

We note that if $I_\varepsilon(X_n)=0$ for all $n=1,..,N$ in (\ref{eq:sto_obj}), the objective equals to 0 constantly. To avoid this situation, we select a sampling distribution which emphasizes the rarer events.

\begin{figure}[b]
      \centering
   \includegraphics[width=\linewidth]{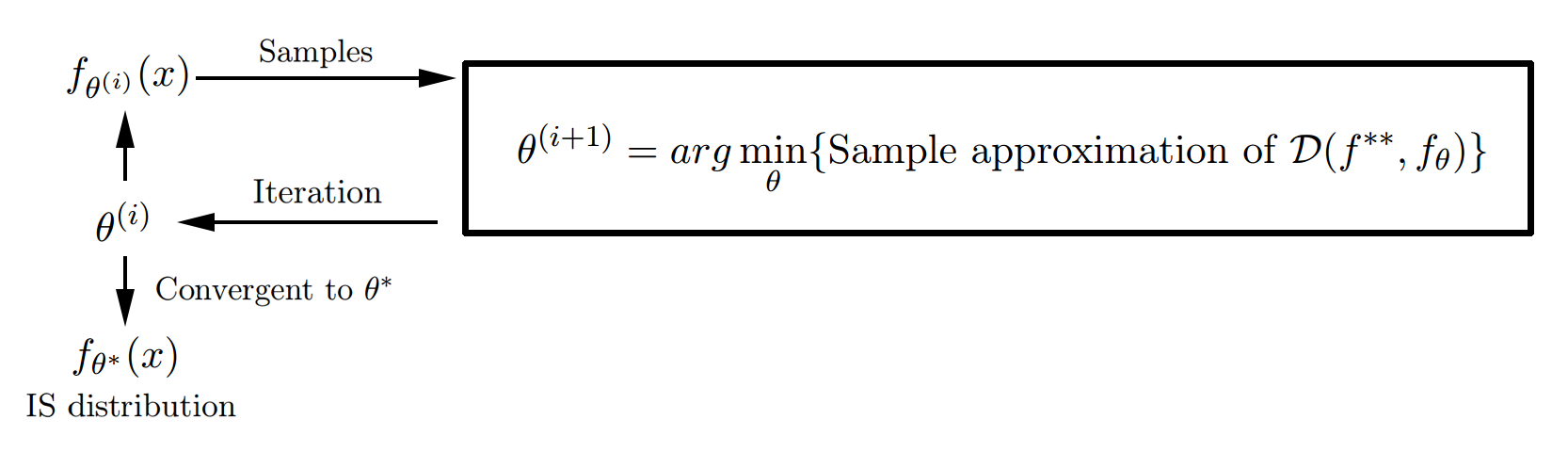}
      \caption{Iterations of Cross Entropy.}
      \label{fig:ce_figure}
\end{figure}

Fig. \ref{fig:ce_figure} shows the iteration procedure of the Cross Entropy method. The core part of the Cross Entropy method is to use the optimizer of the objective function (\ref{eq:sto_obj}) in the $i$th iteration, $\theta^*_i$, as the parameters for the sampling distribution in the next iteration. The underlying idea is that the IS distribution in distribution family $f_\theta$ should better approach the optimal IS distribution. Therefore, as we iterate, we obtain more ``critical'' rare events and have a better estimation of the optimizer which leads to even more ``critical'' rare events in the next iteration. We define the stopping criterion regarding the parameter or the objective value. In practice, we want to start with an appropriate sampling distribution to get a good solution with less iteration. See section \ref{sec:ce_on_rare} for a discussion of initializing a sampling distribution.

We note that if we have two independent variables where $f(x,y)=f(x)f(y)$, we can take a parametric distribution for each variable and have $f_\Theta(x,y)=f_{\theta_1}(x)f_{\theta_2}(y)$, where $\Theta=\{\theta_1,\theta_2\}$. The objective function corresponding to (\ref{eq:sto_obj}) is \begin{multline}
\label{eq:ind_ce}
\max_\theta \ \frac{1}{N} \sum_{n=1}^{N} I_\varepsilon(X_n,Y_n) \frac{f(X_n,Y_n)}{f_{\Theta_s}(X_n,Y_n)} (\ln f_{\theta_1}(X_n)+\\ \ln f_{\theta_2}(Y_n)),
\end{multline}
which can be decoupled into two optimization problem over $\theta_1$ and $\theta_2$ respectively and $ I_\varepsilon(X_n,Y_n) \frac{f(X_n,Y_n)}{f_{\Theta_s}(X_n,Y_n)}$ is a known constant given $\{X_n,Y_n\}$. 

We implement the Cross Entropy on the Piecewise Mixture Distribution with one variable. We note that we can apply the results to the lane change model, since the Cross Entropy objective function of independent variables can be implemented in (\ref{eq:ind_ce}).

\subsection{Optimization Function for Piecewise Mixture Distributions}

We propose a parametric family of IS distribution for Piecewise Mixture Distribution\begin{equation}
{f}_{\theta}(x)=\sum_{i=1}^{k} \tilde{\pi}_i \exp( \theta_{i} x - \kappa (\theta_i))f_i(x |\gamma_{i-1} \leq x< \gamma_i),
\end{equation} where we use exponential change of measure for each piece of distribution and adjust the proportion parameter to $\tilde{\pi}_i$. The parameter is $\theta=\{\theta_1,...,\theta_k,\tilde{\pi}_1,...,\tilde{\pi}_k\}$.

In (\ref{eq:sto_obj}), $c_n=I_\varepsilon(X_n) \frac{f(X_n)}{f_{\theta_s}(X_n)}$ is a known constant given the data, so we simplify the function as \begin{equation}
	\max_\theta \ \frac{1}{N} \sum_{n=1}^{N} c_n \ln f_\theta(X_n).
\end{equation}
We split the samples into index sets $S_i=\{j|\gamma_{i-1} \leq X_j<\gamma_i\}$ for $i=1,...,k$ for each bounded segment. Since $f_i(X_n | \gamma_{i-1} \leq x< \gamma_i) \neq 0$ only if $n\in S_i$, for each $\theta_i$ and $\tilde{\pi}_i$, the optimization function is equivalent to
\begin{multline}
\max_{\theta_i,\tilde{\pi}_i} \frac{1}{N} \sum_{n\in S_i} c_n \ln ( \tilde{\pi}_i \exp( \theta_i X_n - \kappa (\theta_i))\\
f_i(X_n | x< \gamma_{i-1} \leq x< \gamma_i) ) .
\end{multline}

We can further rewrite the optimization function regarding $\theta_i$ and $\tilde{\pi}_i$ respectively. For $\tilde{\pi}_i$, we have\begin{equation}
\max_{\tilde{\pi}_i }\frac{1}{N} \sum_{n\in S_i} c_n \ln \tilde{\pi_i},
\end{equation}
which obtains an analytical form for the optimizer\begin{equation}
\label{eq:pi_update}
\tilde{\pi}_i =\frac{\sum_{n \in S_i} c_n {\bf 1} \{n\in S_i\}}{\sum_{n \in S_i}c_n}.
\end{equation}

For $\theta_i$, we have \begin{multline}
\max_{\theta_i} \frac{1}{N}  \sum_{n\in S_i} c_n \ln \exp( \theta_i X_n - \kappa (\theta_i))  \\f_i(X_n|\gamma_{i-1} \leq x< \gamma_i),
\end{multline} 
which is an exponential change of measure with $D_i$ only.
We note that we can simplify this optimization function by rewriting the log term as
\begin{multline}
\max_{\theta_i} \frac{1}{N}  \sum_{n\in S_i} c_n (\ln \exp( \theta_i X_n - \kappa (\theta_i)) +\\ \ln f_i(X_n |  \gamma_{i-1} \leq x< \gamma_i)),
\end{multline} 
which is equivalent to
\begin{equation}
\max_{\theta_i} \frac{1}{N}  \sum_{n\in S_i} c_n(  \theta_i X_n - \kappa (\theta_i)),
\end{equation}
since the latter term does not depend on $\theta_i$. 

For a bounded exponential distribution with parameter $\lambda$, the Cross Entropy iteration solves\begin{multline}
\max_{\theta_i}\frac{1}{N}  \sum_{n\in S_i} c_n(  \theta_i X_n -\\ \ln\frac{e^{-(\lambda-\theta_i)\gamma_{i-1}}-e^{-(\lambda-\theta_i)\gamma_{i}}}{\lambda-\theta_i}).
\end{multline}

For a bounded normal distribution with parameters $\mu=0$ and $\sigma$, the optimization function for the Cross Entropy iteration is 
\begin{multline}
\max_{\theta_i} \sum_{n\in S_i} c_n X_n \theta_i-(\sum_{n\in S_i} c_n )(\frac{\sigma^2 \theta_i^2}{2} +\\
\ln\frac{\Phi(\frac{\gamma_{i}-\theta_i \sigma^2}{\sigma})-\Phi(\frac{\gamma_{i-1}-\theta_i \sigma^2}{\sigma})}{\Phi(\frac{\gamma_{i}}{\sigma})-\Phi(\frac{\gamma_{i-1}}{\sigma})}).  
\end{multline}

\subsection{Discussion on Numerical Implementation}
\label{sec:tip_prac}
We have presented the optimization functions for Cross Entropy iterations, but we cannot reliably apply these equations in practice without considering some of the problematical numerical details. In this section, we discuss methods to overcome these numerical issues.
\subsubsection{Initializing Cross Entropy Iterations for Rare Events}
\label{sec:ce_on_rare}
Since rare events occur with small probability, using the original distribution as sampling distribution to start the Cross Entropy iterations it becomes computationally burdensome to sample a single rare event. One possible approach is to initialize with guess of sampling distribution. When we have some rough knowledge about the optimal IS distribution, we can use the knowledge to construct a proper sampling distribution. 
 
For cases where we have little knowledge about the optimal IS distribution, we construct adaptive events that gradually reduce the rarity. For rare events denoted by $\varepsilon$, we define the sequence of events to be $\varepsilon_1 \supset \varepsilon_2 \supset ... \supset \varepsilon_n \supset \varepsilon$, where $\varepsilon_1$ is not rare for our initializing sampling density. For each iteration $t$, we gradually reduce the rare event set $\varepsilon_t$ and use $\varepsilon_t$ to replace $\varepsilon$ in the objective function. Since $\varepsilon_t$ is a subset of $\varepsilon_{t-1}$, the IS distribution for $\varepsilon_{t-1}$ also provides more chances for samples from $\varepsilon_t$. We use the optimal solution in $(t-1)$th iteration $\theta^*_{t-1}$ as the sampling parameter $\theta_t$ for the next iteration and choose $\varepsilon_t$ to have a relatively larger probability to occur under $f_{\theta_t}$. Since $\varepsilon_t$ gradually approaches $\varepsilon$ as we iterate, eventually we obtain the optimal parameters for $\varepsilon$.

\subsubsection{Adjusting sample size $N$}

The choice of sample size $N$ should not only depend on the total number of rare events obtained in each iteration. For each parameter of interest, we need sufficient non-zero $c_n$'s to guarantee the qualification of the estimation. We note that the parameters estimation depend only on the rare event in the corresponding piece, so we adjust sample size $N$ to ensure that each piece with large portion $\tilde{\pi}_i$ contains enough rare event samples.

\subsubsection{Setting a lower bound for $\tilde{\pi}_i$}

When we update $\tilde{\pi}_i$ in (\ref{eq:pi_update}), if $c_n=0$ for all $n \in S_i$, meaning that there is no rare event sample in the piece, we have $\tilde{\pi}_i=0$. When we have $\tilde{\pi}_i=0$, the support of the IS distribution will differ from the original distribution. We note that it might cause bias in our simulation analysis. On the other hand, once $\tilde{\pi}_i$ hits 0, it will be 0 in the following iterations. Therefore, we need to keep $\tilde{\pi}_i>0$. Setting a low bound for $\tilde{\pi}_i$, for example, 0.01, when there is no rare event for piece $i$, gives an efficient IS distribution while avoiding the problems.

\subsubsection{Updating parameter $\theta_i$}

The absence of rare event samples also leads to failures in updating $\theta_i$. In this case, we use either the value of $\theta_i$ in the last iteration, or we set it to 0, i.e. reset the distribution as the real distribution. We note that we can tolerant some inaccurate estimation if $\tilde{\pi}_i$ is small, since a small $\tilde{\pi}_i$ indicates that this piece might not be important to the rare events.

\subsubsection{Changing truncation $\gamma_i$}
\label{sec:heuristic}
The truncations of the Piecewise Mixture Distribution are fixed throughout the Cross Entropy method. Thus, if there is a bad selection of truncation in our original distribution model, the Cross Entropy cannot give an efficient IS distribution. The changing of truncation points is hard to implement by optimization, so we use a heuristic approach for adjusting the truncation points to emphasize the tail part of the Piecewise IS distribution.

In any iteration, if the number of rare events is not enough to properly update the parameters, we check $\tilde{\pi}_i$ of the current sampling distribution. If the $\tilde{\pi}_k$ of the tail piece is the largest possible value, we increase the value of the all truncation points except $\gamma_0$ with a certain value. Shifting the truncation gives more weight to the tail part. Then by sampling from the adjusted distribution, we check if the number of events of interest is sufficient. We repeat these actions until we obtain enough rare events in the iteration. 

We propose this heuristic approach, since the flexibility of the Piecewise Mixture Distribution is not fully exploited if we cannot change the truncation points. We note that finding a more systematic procedure to locate the knots remains an open question.

\section{Simulation Analysis}
\label{sec_simulation}
\subsection{Automated Vehicle Model}

First, we present our Piecewise Mixture Models for $R^{-1}$ and $TTC^{-1}$ and then compare the results with the single parametric distribution model used in \cite{Zhao2016AcceleratedTechniques}. For both approaches, we divide the data of $TTC^{-1}$ into three segments regarding the range of $v$. Since the three segments are similar in distribution, we only show the results of the segment for $v$ in the range of 15 to 25 m/s.

\subsubsection{Piecewise mixture models for $R^{-1}$ and $TTC^{-1}$}

\begin{figure}[t]
      \centering
   \includegraphics[width=\linewidth]{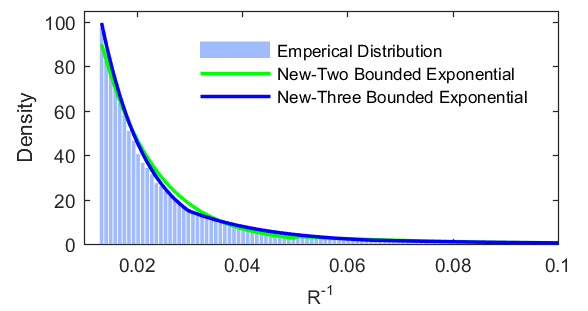}
      \caption{Piecewise Mixture Distribution fitting for $R^{-1}$.}
      \label{fig:r_new}
\end{figure}

Fig. \ref{fig:r_new} shows the fitting of $R^{-1}$ using two bounded exponential distributions and three bounded exponential distributions. Adding one more truncation point provides a better fitting to the body part of distribution while having the same fitting of tail.

\begin{figure}[b]
      \centering
   \includegraphics[width=\linewidth]{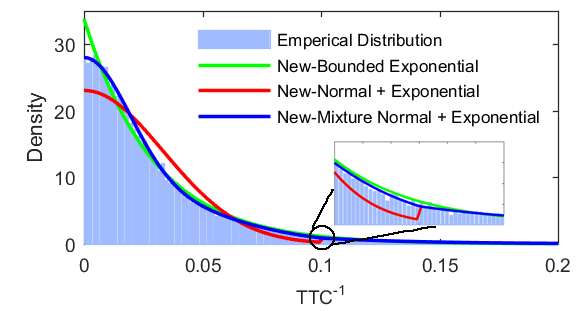}
      \caption{Piecewise Mixture Distribution fitting for $TTC^{-1}$ given $v_L$ between 15 and 25 m/s.}
      \label{fig:ttc2_new}
\end{figure}

In Fig. \ref{fig:ttc2_new}, we truncated the data into two parts. For the tail part, we use the exponential distribution. For the body part, the mixture of two normal distributions gives a better fit. The Piecewise Mixture Models enable us to use different distributions for the body part and the tail part.

\subsubsection{Comparison with single parametric distribution models}

\begin{figure}[thpb]
      \centering
   \includegraphics[width=\linewidth]{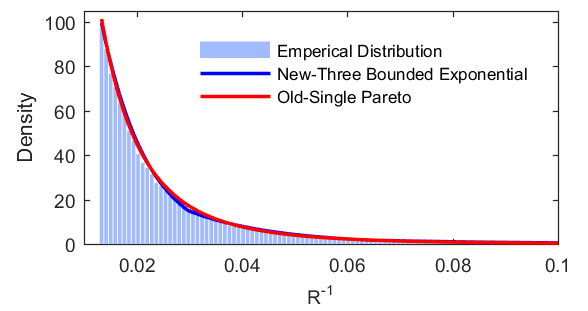}
		\caption{Comparison of fitting for $R^{-1}$.}
      \label{fig:r_old}
\end{figure}

\begin{figure}[thpb]
      \centering
   \includegraphics[width=\linewidth]{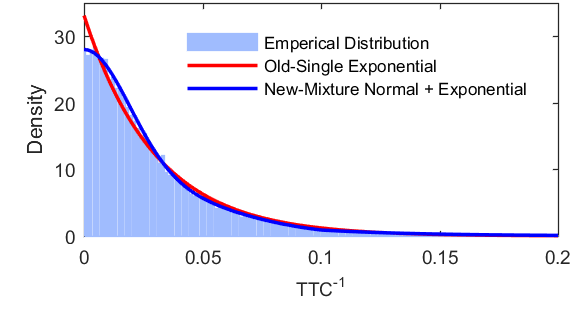}
      \caption{Comparison of fitting for $TTC^{-1}$ given $v_L$ between 15 and 25 m/s.}
      \label{fig:ttc2_old}
\end{figure}

Fig.\ref{fig:r_old} and Fig.\ref{fig:ttc2_old} compare the new model and the previous model. We note that Piecewise Mixture Models provide more flexibility in data fitting. 

\subsection{Cross Entropy Results}

Here, we use the lane change model to exemplify the Cross Entropy method. For the three variables $R, TTC, v$, the distribution is $f(R,TTC,v)=f(v)f(R)f(TTC|v)$ where $f(v)$ is the empirical distribution. Since we have three conditional distributions of $TTC$ regarding the value of $v$, we find the IS distributions independently for each case. We present the results for $v$ from 5 to 15 m/s.

\begin{figure}[t]
      \centering
   \includegraphics[width=\linewidth]{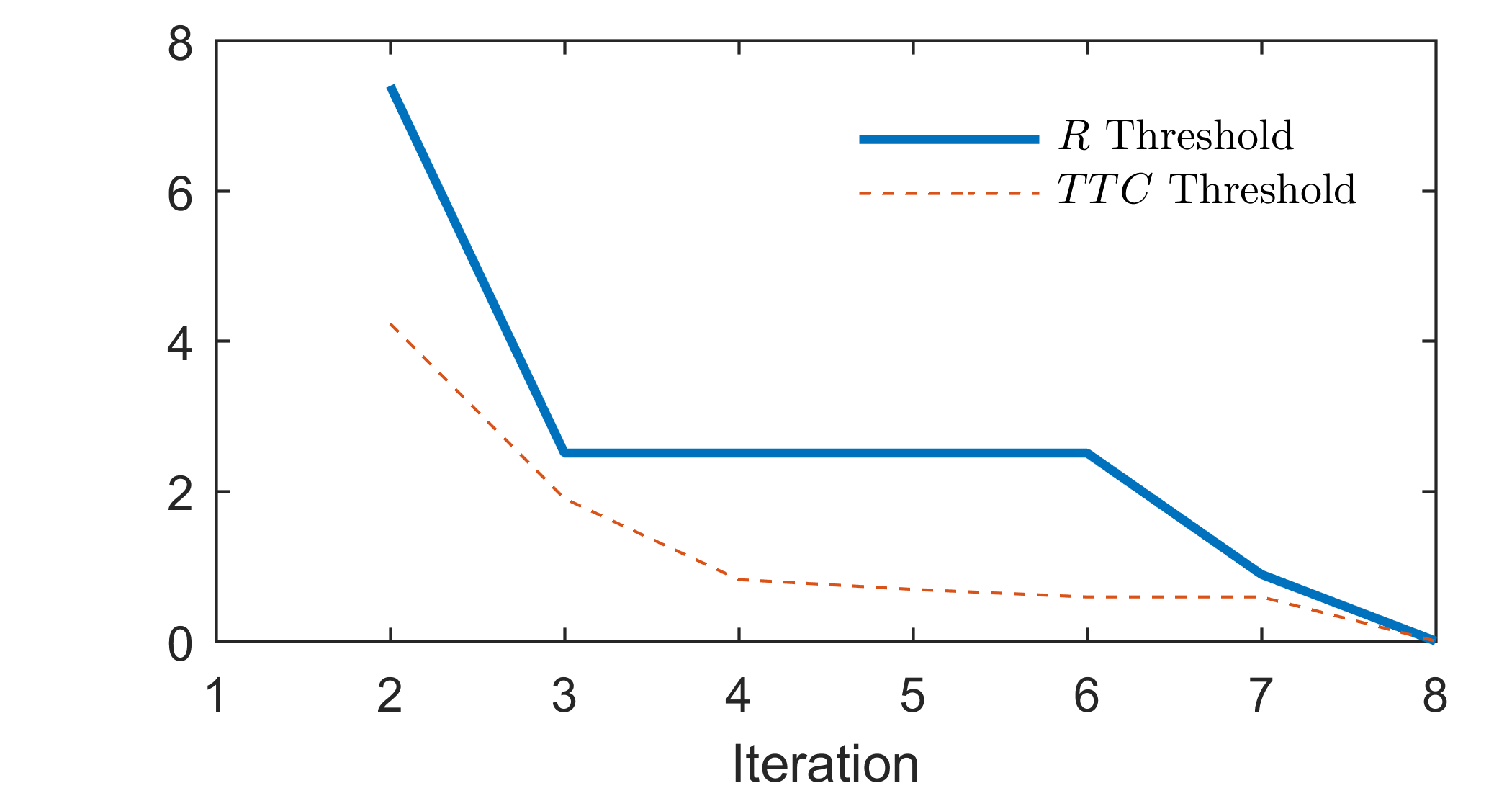}
		\caption{Cross Entropy iterations with sequence of events with thresholds for crash. We leave iteration 1 blank to keep the x-axis consistent with Fig. \ref{fig:ce_r_seg1_mv} and \ref{fig:ce_ttc_seg1_mv}.}
      \label{fig:ce_thrs_seg1_mv}
\end{figure}

We assume that we have less information about the relation between the distribution of variables and the rare events. Our objective is to construct adaptive rare events to help us approach the IS distribution. We recall that our original lane change model determines whether a crash happens by checking to see if the value of $R$, the range between two vehicles, reaches 0. Meanwhile, the $TTC$ also goes to 0 when a crash happens. To construct events less rare than a crash, we relax the criterion for crash to be either $R$ hits $t_{R}>0$ or $TTC$ hits $t_{TTC}>0$. By changing these two thresholds, $t_{R}$ and $t_{TTC}$ as shown in Fig. \ref{fig:ce_thrs_seg1_mv}, we construct the adaptive rare events sequence for the Cross Entropy iterations. We use sample size $N=1000$ for each iteration. 

\begin{figure}[b]
      \centering
   \includegraphics[width=\linewidth]{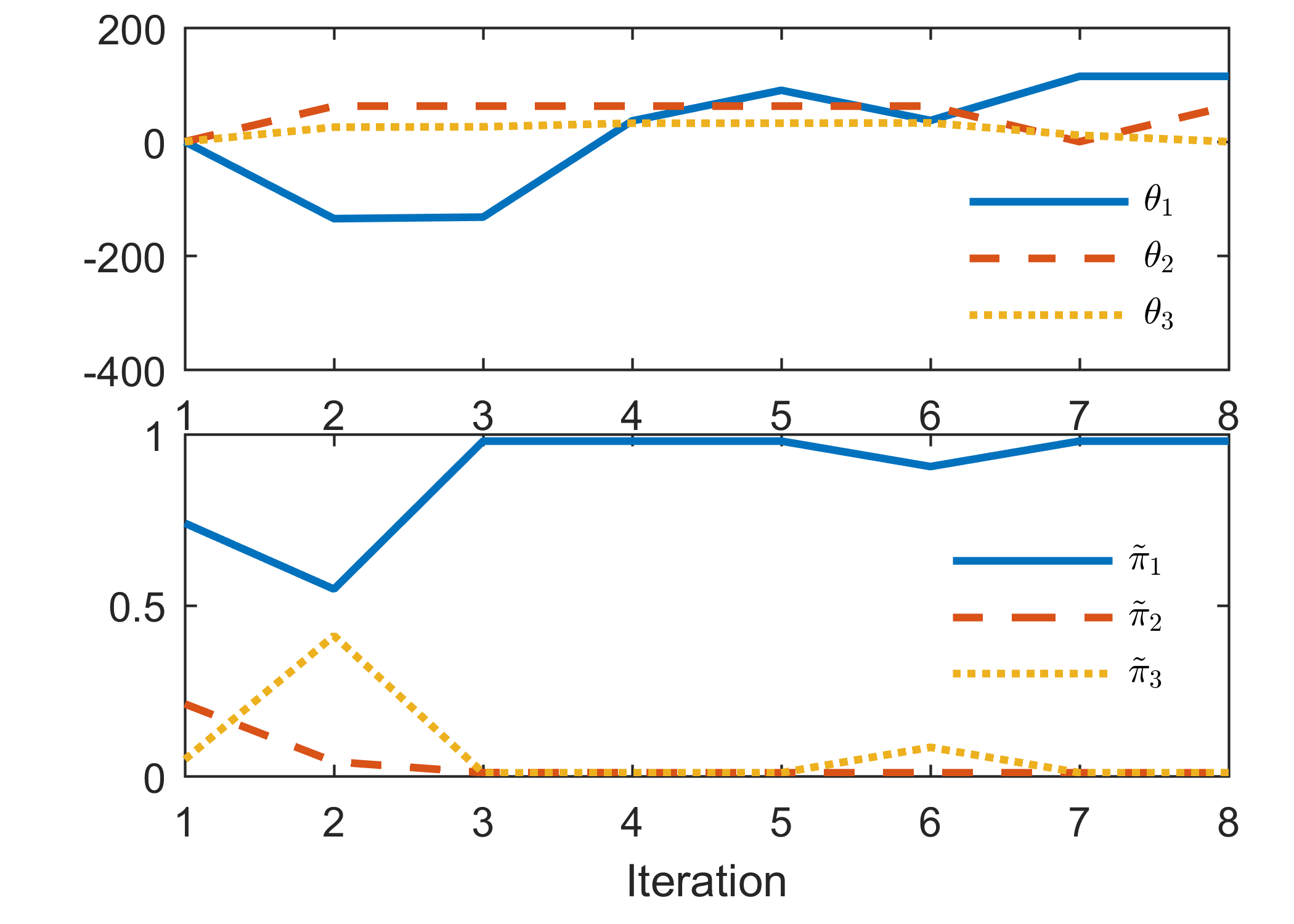}
		\caption{Cross Entropy iterations with sequence of events of $R^{-1}$ for $v$ from 5 to 15 m/s.}
      \label{fig:ce_r_seg1_mv}
\end{figure}

\begin{figure}[thpb]
      \centering
   \includegraphics[width=\linewidth]{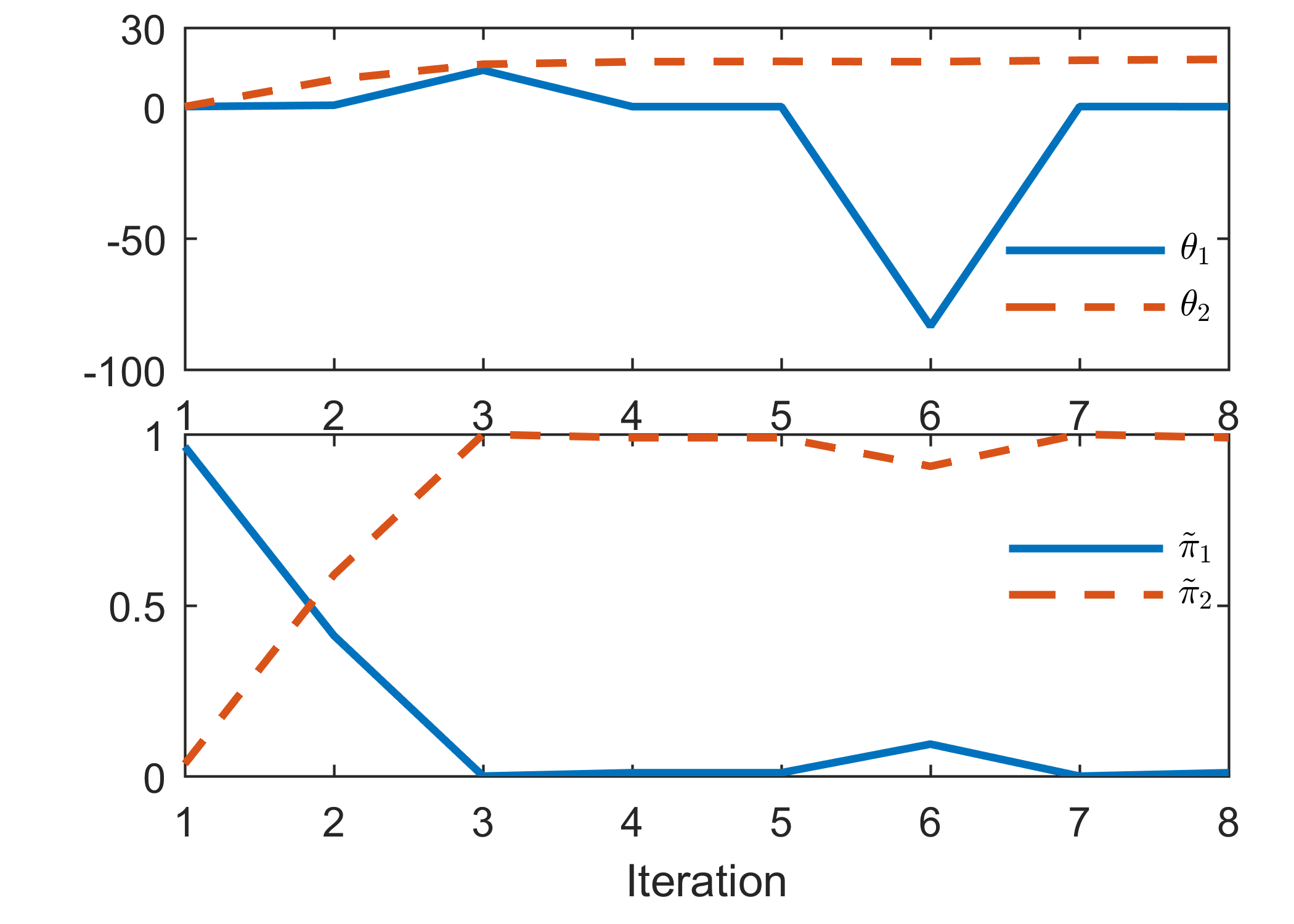}
		\caption{Cross Entropy iterations with sequence of events of $TTC^{-1}$ for $v$ from 5 to 15 m/s.}
      \label{fig:ce_ttc_seg1_mv}
\end{figure}

\begin{figure}[thpb]
      \centering
   \includegraphics[width=\linewidth]{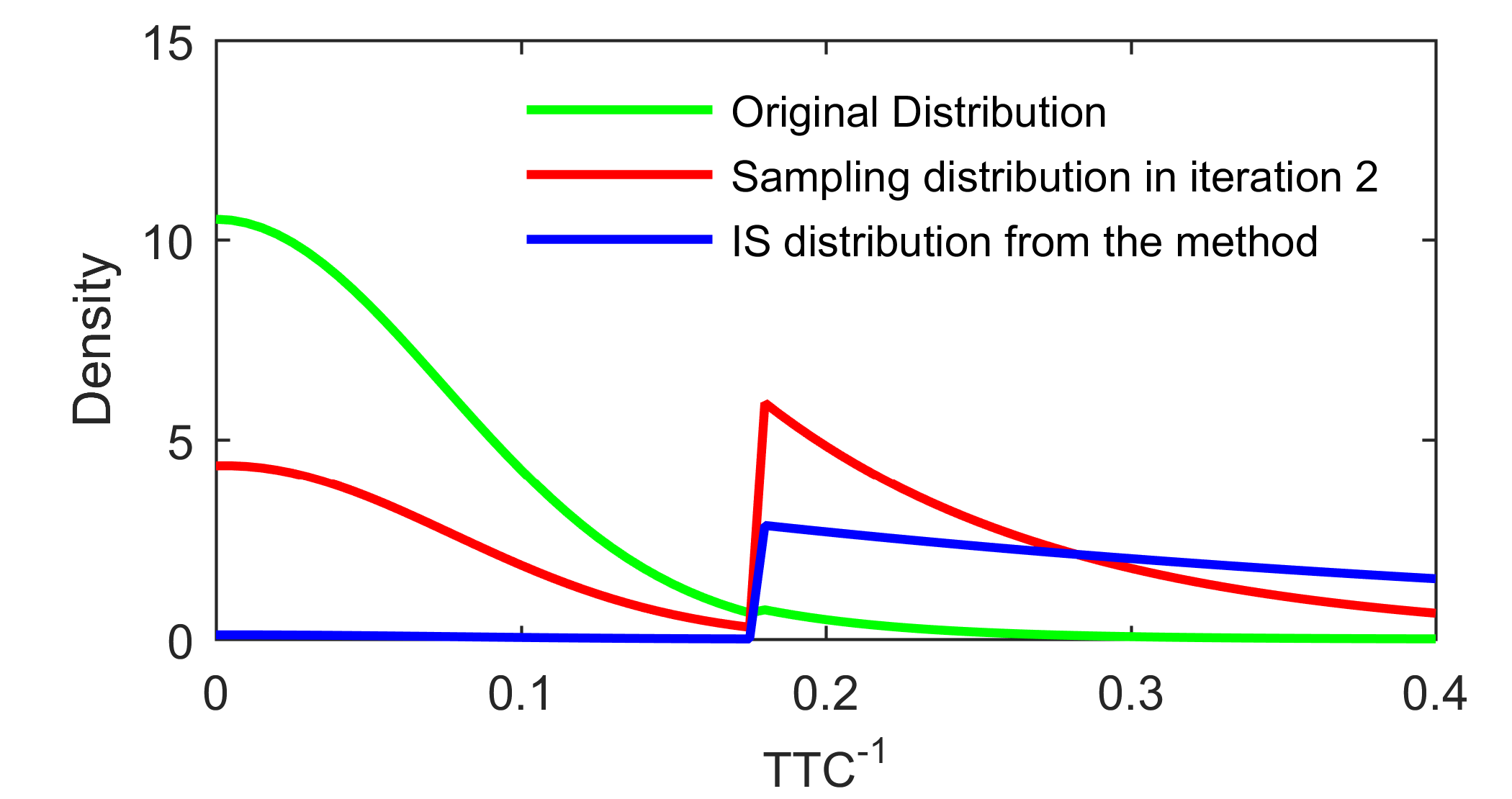}
		\caption{Distribution change through Cross Entropy iterations with sequence of events of $TTC^{-1}$ for $v$ from 5 to 15 m/s.}
      \label{fig:ce_ttc_dist_change_mv}
\end{figure}

Fig. \ref{fig:ce_r_seg1_mv} and \ref{fig:ce_ttc_seg1_mv} show the parameters present in each of the iterations. We observe that the parameters stabilize gradually. Fig. \ref{fig:ce_ttc_dist_change_mv} shows how the distribution changes gradually from the original distribution to the IS distribution. We note that the density moves toward the tail part as we iterate.

\subsection{Simulation Results}

In our simulation experiments, we set the convergence criterion as the relative half-width of $100(1-\alpha)\%$ confidence interval drops below $\beta$. In this case, we use $\alpha=0.2$ and $\beta=0.2$ to study the number of samples needed for convergence. Our goal is to compare the efficiency of the Piecewise Mixture Distribution and single exponential distribution models.

\begin{figure}[b]
      \centering
   \includegraphics[width=\linewidth]{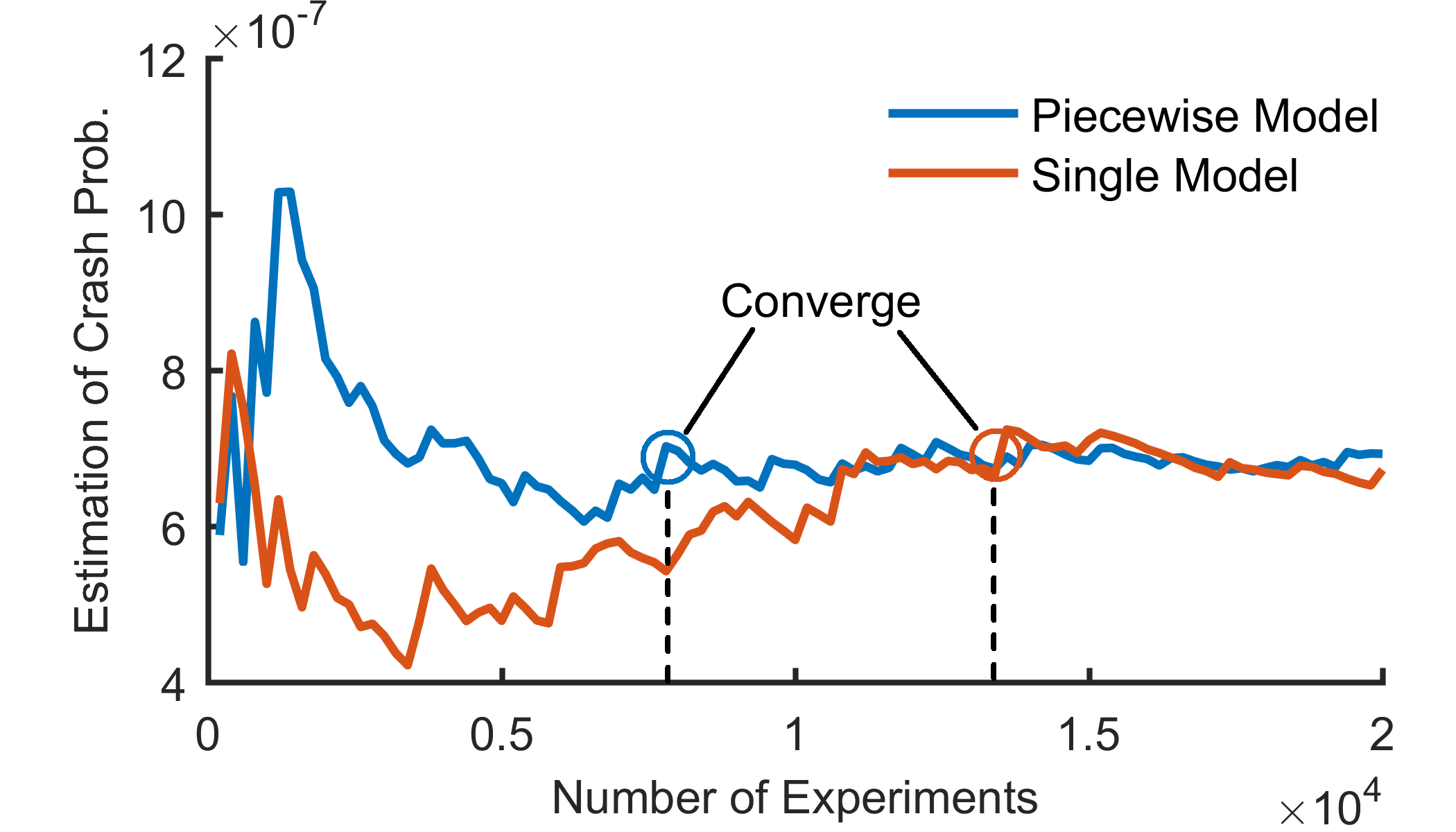}
      \caption{Estimation of crash probability for one lane change using piecewise and single accelerated distributions.}
      \label{fig:result_estimation}
\end{figure}


Fig. \ref{fig:result_estimation} shows that both models give a similar estimation as the number of experiments grows large, and that the Piecewise Mixture Distribution model converges slightly faster than the single parametric model. The circles show that the relative half-width of the Piecewise Mixture Distribution model reaches the target confidence value after 7800 samples, whereas the single parametric model needs about 13800 samples. Using the Piecewise Mixture Distribution model reduced the sample size by 44\%.

To reduce stochastic uncertainty, we repeat the tests 10 times and calculate the average. It takes 7840 samples on average to obtain a converged estimation using the Piecewise Mixture Distribution model, whereas it takes 12320 samples on average using the single accelerated distribution model to converge. Table \ref{table:n_ratio} compares the two models with the crude Monte Carlo method \cite{Asmussen2007StochasticAnalysis}. We estimate the number needed for convergence of crude Monte Carlo by using the fact that the number of events of interest occurring is Binomial distributed. We compute the standard deviation of the crude Monte Carlo estimation $\hat{P}(x \in \varepsilon)$ by \begin{equation}
std(\hat{P}(x \in \varepsilon))=\sqrt{\frac{\hat{P}(x \in \varepsilon)(1-\hat{P}(x \in \varepsilon))}{n}},
\end{equation} 
which allows us to estimate \begin{equation}
\hat{N}=\frac{ z_{\alpha/2}^2 (1-\hat{P}(x \in \varepsilon))}{ \beta^2  \hat{P}(x \in \varepsilon)}, 
\end{equation}
where $z_{\alpha/2}$ is the $(1-\alpha/2)$ quantile of normal distribution. We calculate the required sample size $N$ of crude Monte Carlo in Table \ref{table:n_ratio} from an estimation $\hat{P}(x \in \varepsilon)=7.4\times 10^{-7}$ with $80 \%$ confidence interval $(7.0\times 10^{-7},7.8\times 10^{-7})$.
\begin{table}[t]
\centering
\caption{Number of samples (N) needed to converge. }
\label{table:n_ratio}
\begin{tabular}{l|lll}
\hline
\hline
           & Piecewise   & Single & Crude     \\ \hline
N           & 7840 & 12320  & $5.5\times 10^7$ \\ \hline
Ratio to Piecewise & 1    & 1.57   & $7\times 10^3$   \\ \hline \hline
\end{tabular}
\end{table}

\begin{figure}[t]
      \centering
   \includegraphics[width=\linewidth]{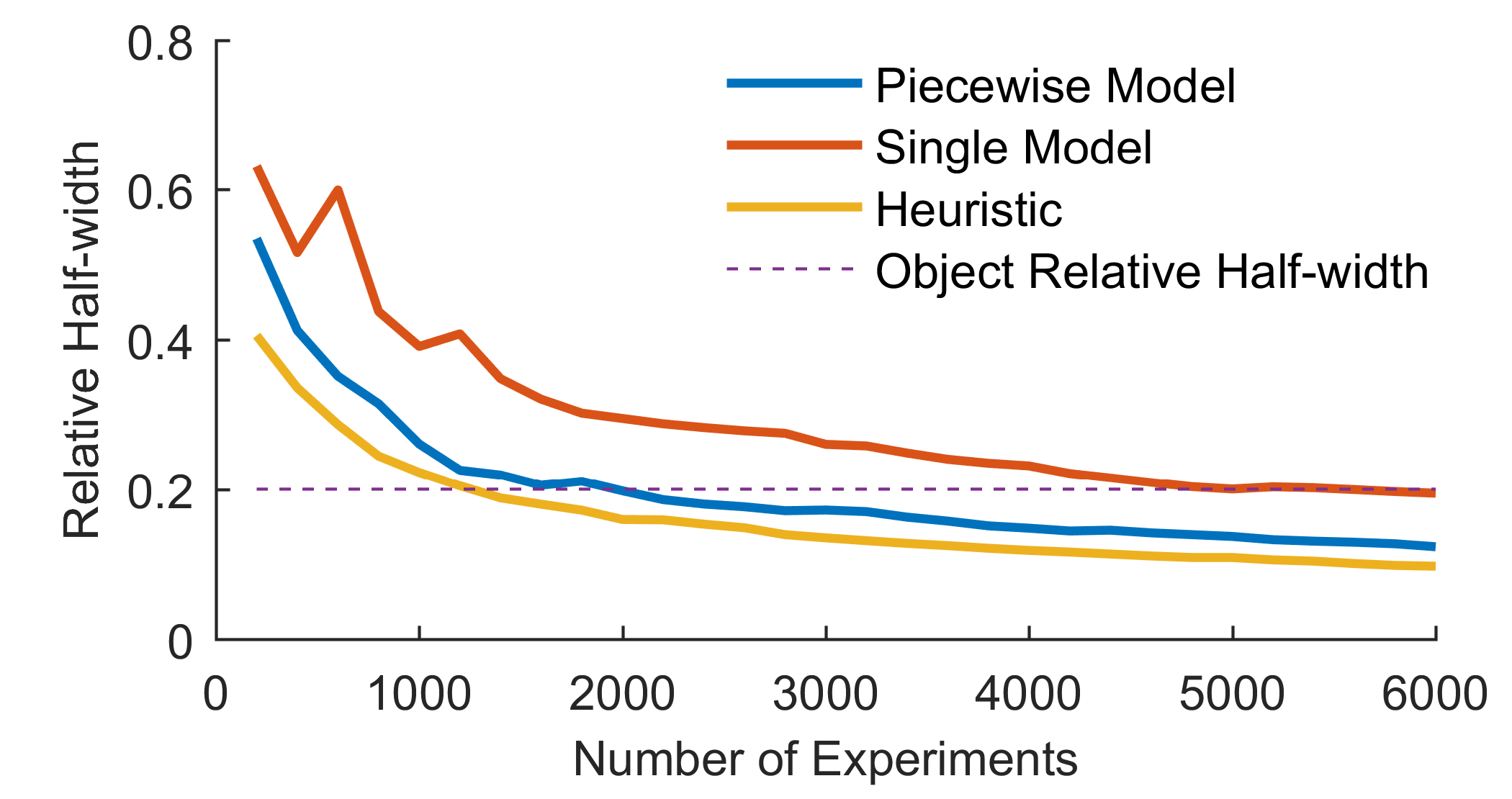}
      \caption{Relative half-width of crash probability estimation for one lane change with leading vehicle's speed in range of 5 to 15 m/s, comparing single, piecewise and heuristic accelerated distributions.}
      \label{fig:seg1_simulation}
\end{figure}

Finally, we apply the heuristic approach in Section \ref{sec:heuristic} to the data segment with $v$ from 5 to 15 m/s. We run simulations with this segment and compare the results with the standard approach for the ewise Mixture Distribution and single parametric distribution models. Fig. \ref{fig:seg1_simulation} shows the convergence of confidence half-width. We note that the relative half-width of the heuristic, which is smaller than the standard approach for the Piecewise Mixture Distribution model, indicates that the latter model's performance can be further improved.

\section{Conclusions}
\label{sec:conclusions}
This paper proposed a new model for accelerated evaluation of AVs. The Piecewise Mixture Distribution Models provide more accurate fitting to the surrounding human-controlled vehicle behaviors than the single parametric model used in the literature. The proposed model was more efficient and reduced the evaluation time by almost half than single parametric model. The Cross Entropy procedure described in this paper effectively worked in this scenario analysis. We provided practical solutions to deal with the numerical issues which occurred while calculating the optimal parameters. The heuristic approach exploited the flexibility of the Piecewise Mixture Distribution structure. Testing the proposed model on a large dataset of cut-in crashes caused by improper lane changes, the Piecewise Mixture Distribution model reduced the simulation cases by about 33\% compared with the single parametric model under the same convergence requirement. Moreover, the proposed model was 7000 times faster than the Crude Monte Carlo method. 

Table \ref{table:summarize} summarizes the comparison of the computation efforts between the models. We note that using the Piecewise Mixture Distribution model increases the number of parameters estimated, where the estimation of parameters is almost instant. In the Cross Entropy stage, the number of simulations required for the Piecewise model is not significantly less than the single parametric model, because we assume no knowledge about the optimal IS distribution for the Piecewise model. Overall, the Piecewise model needs fewer simulations to reach the same confidence level compared to the single parametric Model.

\begin{table}[t]
\centering
\caption{Comparison of the computation time between single parametric model and piecewise model.}
\label{table:summarize}
\begin{tabular}{l|l|l|l}
\hline
\hline
Stages        & Crude & Single                                                                                    & Piecewise                                                                               \\ \hline
Fitting       & -     & \begin{tabular}[c]{@{}l@{}}4 parameters to\\   estimate\end{tabular}                      & \begin{tabular}[c]{@{}l@{}}18 parameters to\\   estimate\end{tabular}                   \\ \hline
Cross Entropy & -     & \begin{tabular}[c]{@{}l@{}}30,000 simulations\\   4 parameters\end{tabular} & \begin{tabular}[c]{@{}l@{}}24,000 simulations \\ 18 parameters\end{tabular} \\ \hline
Simulation &  \begin{tabular}[c]{@{}l@{}}$5.5 \times 10^7$ \\simulations  \end{tabular}   & \begin{tabular}[c]{@{}l@{}}12,320 \\simulations  \end{tabular} & \begin{tabular}[c]{@{}l@{}}7840 \\simulations  \end{tabular} \\ \hline \hline
\end{tabular}
\end{table}

\appendices

\section{Inverse CDF of Piecewise Mixture Distributions}
	\label{apped_inv}
	We can sample from Piecewise Mixture Distribution by the inverse CDF approach. Here, we derive the inverse CDF for Piecewise Mixture Distribution. 
	
	The CDF of Piecewise Mixture Distribution (\ref{eq:CDF}) can split into\begin{equation}
    \resizebox{\hsize}{!}{$
F(x)= \begin{cases} 
	... \\
	\sum_{j=1}^{i-1}\pi_j + \pi_i F_i(x | \gamma_{i-1} \leq x< \gamma_i) &  \gamma_{i-1} \leq x< \gamma_i\\
	...
	\end{cases}.$}
\end{equation}
Therefore the inverse function can be written as\begin{equation}
\label{eq:icdf}
\resizebox{\hsize}{!}{$F^{-1}(y)=\begin{cases} 
	... \\
	F_i^{-1}(\frac{y-\sum_{j=1}^{i-1}\pi_j}{\pi_i} | \gamma_{i-1} \leq x< \gamma_i) & \sum_{j=1}^{i-1}\pi_j \leq y < \sum_{j=1}^{i}\pi_j\\
	...
	\end{cases}.$   }
\end{equation}
where $F_i^{-1}$ is the inverse conditional CDF of $F_i$. Below, we give two example of inverse conditional CDF.
	
For the inverse CDF of conditional exponential distribution, we have
\begin{multline}
F_\theta^{-1}(y|F_\theta(\gamma_1)\leq y<F_\theta(\gamma_2))=\\F_\theta^{-1}((F_\theta(\gamma_2)-F_\theta(\gamma_1))y+F_\theta(\gamma_1)),
\end{multline}
where $F$ and $F^{-1}$ are the CDF and inverse CDF of exponential distribution.

	For conditional normal distribution, the inverse CDF is\begin{multline}
F_\theta^{-1}(y|F_\theta(\gamma_1) \leq y < F_\theta(\gamma_2)) =\\ \sigma \Phi^{-1} ((\Phi(\frac{\gamma_2-\theta \sigma^2}{\sigma})-\Phi(\frac{\gamma_1-\theta \sigma^2}{\sigma}))y  + \\ \Phi(\frac{\gamma_1-\theta \sigma^2}{\sigma}))+\theta \sigma^2.
\end{multline}
\\    
\section{EM Algorithm for Mixture Bounded Normal Distribution}
\label{apped_em}
Here, we present a numerical MLE algorithm with mixture bounded normal distribution. The steps are as follows.

{\bf ALGORITHM:}
%
%
%
\begin{enumerate}
	\item Initialize $\{p_j,\sigma_j\}$, $j=1,...,m.$
	\item E step: update \begin{equation}
\tau_n^j= \frac{p_j f_j(X_n|\sigma_j)}{\sum_{j=1}^{m}p_j f_j(X_n|\sigma_j)}.
\end{equation} 
	\item M step: update \begin{equation}
	p_j=\frac{\sum_{n=1}^{N} \tau_n^j }{N}
\end{equation}
    and 
	\begin{equation}
	\sigma_j=\arg \min_{\sigma_j} - \tau_n^j \ln \frac{\sigma_j (\Phi(\frac{\gamma_2}{\sigma_j})-\Phi(\frac{\gamma_1}{\sigma_j}))}{\phi(\frac{X_n}{\sigma_j})}.
\end{equation}

	\item Repeat 2 and 3 until $\mathcal{L}(\theta|D)$ converges.
\end{enumerate}

\section{Vanilla Cross Entropy Method}
\label{apped_ce}

{\bf ALGORITHM:}\cite{Boer2005AMethod}

\begin{enumerate}
	\item Initialize $\theta_s$.
	\item Sample $\{X_1,...,X_N\}$ from $f_{\theta_s}$ and update \begin{equation}
\theta= \arg\max_\theta \ \frac{1}{N} \sum_{i=1}^{N} I_\varepsilon(X_i) \frac{f(X_i)}{f_{\theta_s}(X_i)} \ln f_\theta(X_i).
\end{equation} 
	\item Update \begin{equation}
	\theta_s=\theta.
\end{equation}
    	
	\item Repeat 2 and 3 until $\theta$ ``converges''.
\end{enumerate}


\ifCLASSOPTIONcaptionsoff
  \newpage
\fi

\bibliographystyle{IEEEtran}
\bibliography{Mendeley_MTC_AE.bib}

\begin{IEEEbiography}[{\includegraphics[width=1in,height=1.25in,clip,keepaspectratio]{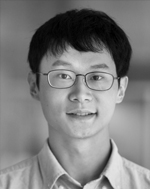}}]{Zhiyuan Huang}
Zhiyuan Huang is a second year pre-candidate Ph.D. student in Industrial and Operations Engineering at the University of Michigan, Ann Arbor. His research interests include simulation and stochastic optimization.
\end{IEEEbiography}

\begin{IEEEbiography}[{\includegraphics[width=1in,height=1.25in,clip,keepaspectratio]{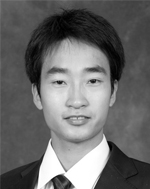}}]{Ding Zhao}
Ding Zhao received the Ph.D. degree in 2016 from the University of Michigan, Ann Arbor. He is currently a Research Fellow in the University of Michigan Transportation Research Institute. His research interest includes evaluation of connected and automated vehicles, vehicle dynamic control, driver behaviors modeling, and big data analysis.
\end{IEEEbiography}

\begin{IEEEbiography}[{\includegraphics[width=1in,height=1.25in,clip,keepaspectratio]{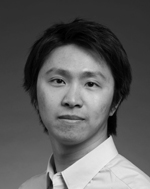}}]{Henry Lam}
Henry Lam received the B.S. degree in actuarial science from the University of Hong Kong in 2005, and the A.M. and Ph.D. degrees in statistics from Harvard University, Cambridge, in 2006 and 2011.

From 2011 to 2014, he was an Assistant Professor in the Department of Mathematics and Statistics at Boston University. Since 2015, he has been an Assistant Professor in the Department of Industrial and Operations Engineering at the University of Michigan, Ann Arbor. His research focuses on stochastic simulation, risk analysis, and simulation optimization. Dr. Lam's works have been funded by National Science Foundation and National Security Agency. He has also received an Honorable Mention Prize in the Institute for Operations Research and Management Sciences (INFORMS) George Nicholson Best Student Paper Award, and Finalist in INFORMS Junior Faculty Interest Group Best Paper Competition.
\end{IEEEbiography}

\begin{IEEEbiography}[{\includegraphics[width=1in,height=1.25in,clip,keepaspectratio]{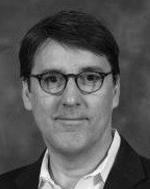}}]{Dave J. LeBlanc}
Dave J. LeBlanc received a Ph.D. in aerospace engineering from the University of Michigan, and master's and bachelor's degrees in mechanical engineering from Purdue University. Dr. David J. LeBlanc is currently an associate research scientist, has been at UMTRI since 1999. Dr. LeBlanc's work focuses on the automatic and human control of motor vehicles, particularly the design and evaluation of driver assistance systems.
\end{IEEEbiography}



\end{document}